# Effect of Structural Transition on Magnetic Properties of Ca and Mn co-substituted BiFeO$_3$ Ceramics

Pawan Kumar and Manoranjan Kar[*]

Department of Physics, Indian Institute of Technology Patna, Patna-800013, India.

* Corresponding author Email: mano@iitp.ac.in, Ph: +91612-2552013, Fax: +91612- 2277383

**Abstract:**

Composition-driven structural transitions in $Bi_{1-x}Ca_xFe_{1-x}Mn_xO_3$ ceramics prepared by the tartaric acid modified sol-gel technique have been studied to analyze its effect on the magnetic properties of bismuth ferrite (BiFeO$_3$). It was observed that the co-substitution of Ca & Mn at Bi & Fe sites in BiFeO$_3$ (BFO) significantly suppress the impurity phases. The quantitative crystallographic phase analysis has been carried out by double phase Rietveld analysis of all the XRD patterns which indicates the existence of compositional driven crystal structure transformation from rhombohederal (*R3c* space group, lower crystal symmetry) to the orthorhombic (*Pbnm* space group, higher crystal symmetry) with the increase in substitution concentration due to excess chemical pressure (lattice strain). Magnetic measurements reveal that co-substituted BFO nanoparticles for x = 0.15 have enhanced remnant magnetization about 14 times that of pure one due to the suppression of cycloid spin structure which could be explained in terms of field induced spin reorientation and weak ferromagnetism. However, at the morphological phase boundary (x = 0.15), the remnant and maximum magnetization at 8 T reaches a maximum which indicates almost broken spin cycloid structure and further increase in substitution results in the reduction of both magnetizations due to the appearance of complete antiferromagnetic ordering in the orthorhombic structure because of the significant contribution from the crystallographic phase of *Pbnm* space group (as obtained from double phase Rietveld analysis).

# 1. Introduction

Multiferroics is an important class of multifunctional materials with coupled electric, magnetic, and structural order parameters. Among all the identified single-phase multiferroics, BFO is an interesting multiferroic material which exhibits the coexistence of ferroelectric order and G-type canted antiferromagnetic order well above room temperature ($T_c$ = 1103 K and $T_N$ = 643 K) [1]. It has possible applications in multistate memory devices, spintronic devices, magnetically modulated transducers, ultrafast optoelectronic devices and sensors [2-5]. The crystal structure for polar phase of BFO at room temperature is described by rhombohederally distorted perovskite with R*3c* space group. This space group allows the antiphase octahedral tilting and ionic displacement from the centrosymmetric position along [111]$_C$ direction of the parent cubic perovskite unit cell. The ferroelectricity in this compound is due to the off-centre structural distortions of cations whereas the magnetism due to local spins. The R3c symmetry allows the existence of weak ferromagnetic moment due to Dzyaloshinky-Moriya interaction but the cycloid spin structure with the periodicity of ~62 nm prevents net magnetization leads to net zero magnetization [6-9]. Recently, it has been reported that the high ferroelectric and ferromagnetic polarization or large magetoelectric coupling constant at room temperature through A- site and/or B-site substitution in BFO [10-15]. Direct evidence of cycloid suppression in A-site substituted samples has been given via nuclear magnetic resonance (NMR) measurements which has been correlated with the structural transition from rhombohedral to orthorhombic crystal structure [10, 14, 15]. This structural phase transition (in morphotropic phase boundary) significantly enhances the magnetization as well as magnetoelectric interaction. However, BFO is generally quite difficult to be prepared in phase pure ceramics or thin film because of its narrow temperature range of phase stabilization. However, several attempts have been made to prepare phase pure by chemical route and the solid state route followed by leaching with nitric acid [16, 17]. The nitric acid leaching is normally used to eliminate above impurity phases which leads to the formation of coarser powders and its poor reproducibility. Hence, we have adopted the chemical route of synthesis for uniform particle size and better reproducibility.

Weak ferromagnetic ordering induced due to Ca substitution at A-site in BFO has been observed by B. Ramchandran et al and D. Khothari et al [18, 19]. The Mn substitution at Fe site of BFO is reported to inhibit the grain growth which resulted in reduced particle size [20] and improve the magnetic as well as electric properties [21, 22]. S. Chauhan et al. had reported the structural phase transition in 15% Mn doped BFO sample due to the distortion in the rhombohedral structure with increasing Mn substitution which resulted in significant enhancement in magnetization [23]. J-Z Huang et al have reported the structural transition and improved ferroelectricity in Ca and Mn co-substituted BFO thin films up to 10% substitution concentration [24].

However, to the best of our knowledge, a systematic study of the influence of crystal structural transition on magnetic properties has not been carried out on Ca and Mn co-substituted BFO ceramics. Due to the fact that co-substitution of Ca and Mn can eliminate the formation of impurity phases and enhance the magnetization of co-substituted BFO ceramics, we have undertaken a study on Ca and Mn co-substituted BFO prepared by the tartaric acid modified sol-gel method and carry out double phase Rietveld analysis to study the crystallographic phases and correlation of different crystal symmetries with the magnetic properties.

## 2. Methods

$Bi_{1-x}Ca_xFe_{1-x}Mn_xO_3$ with x = 0.000, 0.025, 0.005, 0.100, 0.150 and 0.200 (BCFM-0, BCFM-025, BCFM-05, BCFM-10, BCFM-15, BCFM-20 respectively) were prepared by a tartaric acid modified sol-gel technique. The detailed preparation procedure has been discussed in our previous publication [25]. Here, the starting materials as bismuth nitrate, iron nitrate, calcium hydroxide, manganese acetate and tartaric acid (purity $\geq$ 99.0%) were carefully weighted in stoichiometric proportion. The molar ratio of metal nitrates to tartaric acid was taken as 1:2. The resulting material was thoroughly grinded and annealed at 700 $^o$C for 3 hours.

The crystallographic phases of all the samples were determined by the powder X-ray diffraction (XRD) study using 18 kW Cu-rotating anode based Rigaku TTRX III diffractometer, Japan) with CuKα radiation (λ = 1.5418 Å) operating in the Bragg-Brentano geometry in a 2θ range of 10$^o$-120$^o$ at a scan step of 0.01$^o$ (counting time for each step was 2.0 Sec). The microstructural properties of the samples were investigated by using Field Emission Scanning Electron Microscopy (FE-SEM), Hitachi S-4800, JAPAN, operating at an accelerating voltage of 10 kV and equipped with energy-dispersive X-ray spectroscopic (EDS) capability. Room temperature Raman spectra were measured in the backscattering geometry using confocal micro-Raman spectrometer (Seki Technotron Corp., Japan) with the 514.5 nm laser line as excitation source by STR 750 RAMAN Spectrograph using a 100 X microscope. Fourier Transform Infrared Spectra (FT-IR) were recorded at room temperature using the Perkin Elmer (model 400) in the range from 400 to 1200 cm$^{-1}$. The diffuse reflectance spectroscopy has been carried out by using LAMBDA 35 UV-visible spectrophotometer in the range from 200 to 1100 nm. Room temperature magnetization – magnetic field (M-H) measurement for all the samples were carried out by magnetic properties measurement system (MPMS), Quantum Design Inc., USA in the applied magnetic field of maximum ± 80 kOe.

All XRD patterns were analyzed employing Rietveld refinement technique with the help of Fullprof package [26]. The patterns for all the samples could be refined using the *R3c* as well as *Pbnm* space groups. All the occupancy parameters were taken as fixed at their corresponding composition during refinement. Other parameters, such as, Zero correction, scale factor, half width parameters, lattice parameters, atomic fractional position coordinates, thermal parameters were varied during refinement. Background was defined by the sixth order polynomial whereas peak shape by pseudo-Voigt function.

## 3. Results and discussion

The XRD patterns for all the samples annealed at 700 °C have been shown in Fig. 1. The main concern for BFO is Bi vacancies since both anion and cation vacancies can be found in perovskites. However, the formation of $Bi_{1-x}FeO_3$ in our samples is negligible because the phase $Bi_2Fe_4O_9$ is always observed whenever the Bi/Fe ratio falls below one [27]. It is important to note that we do not see any $Bi_2Fe_4O_9$ impurities peaks of significant intensity in our XRD patterns of co-substituted samples. These results confirm the formation of BCFM nanoparticles with negligible impurities. The Goldschmidt tolerance factor (t) in $ABO_3$ ($BiFeO_3$) structure is defined as,

$$t = \frac{(<r_A> + r_o)}{\sqrt{2}(<r_B> + r_o)} \quad \text{------------ (1)}$$

Here, $<r_A>$ and $<r_B>$ are the average radius of A site and B site cations respectively and $r_o$ is the ionic radius of oxygen. The tolerance factor is used to quantify the structural stability of perovskite compounds. When the value of "t" is smaller than one, the compressive strain acts on the Fe-O bonds and hence on Bi-O bonds which induce lattice distortion and leads to the evolution of higher symmetric crystallographic phase such as orthorhombic or tetragonal. Reflections (104) and (110) are clearly separated in the BCFM-05 sample as shown in Fig. 1. On increasing the Ca and Mn content, all the doublets appears merging to give a single peak, which is clearly visible in BCFM-10 and higher substitutions. These results correspond to the lattice distortion in rhombohedral structure of BFO occurring with the increase in the substitution concentration which leads to rhombohedral to orthorhombic crystal phase transition. The crystallite size has been calculated using Scherrer's formula by the Gaussian fits to the observed maximum intensity X-ray diffraction peak [28]. The crystallite size decreases with the increase of substituent ion concentration (mentioned in table 1). This implies the development of lattice strain inside the lattice due to ionic size mismatch between $Ca^{2+}$ (1.06 Å) & $Bi^{3+}$(1.17Å) which lead to local structural disorder and reduces the rate of nucleation as an effect decrease the crystallite size. The crystallite size has been also determined by FE-SEM which shows that the particles have an almost homogeneous distribution as shown in Fig.2 for BCFM-05. The average crystallite size was found to be ~48 nm for BCFM-05 which is close to the value obtained from XRD patterns analysis. The unit cell volume decreases with the increase in the substituent concentration because ionic size of $Ca^{2+}$ is less than that of $Bi^{3+}$. The elemental analysis of samples were carried out using the EDS. The typical EDS pattern of sample BCFM-05 has been shown in Fig. 2. It reveals the presence of Bi, Fe, Ca, Mn and O elements in the sample. No extra peaks have been traced which indicate that there is no contamination in the sample.

The typical Rietveld refinements of XRD patterns for BCFM-15 and BCFM-20 have been shown in Fig. 3. The refined lattice parameters of all the samples have been given in Table 2. The rhombohedral phase was considered for the Rietveld refinement of BCFM-0 XRD profile because it has the characteristic doublet of highest intensity peaks. But the characteristic doublet of highest intensity peaks merges for the co-substituted BFO which clearly indicates the presence of higher crystallographic symmetry. Rigorous fitting with the different structural model/s (*R3c*, *R3c + Pbnm*, *R3c + P4mm*, *R3c + Pm3m*, *R-3c + Pbnm*, etc.) showed that the observed XRD patterns of co-substituted samples are a result of the superposition of two spectral contributions (*R3c + Pbnm*). The respective contribution of both crystallographic phases has been given in Table 2. It is worth noting that we have not got satisfactory results by taking a single space group in Rietveld refinement. Since Raman spectroscopy is sensitive to the atomic displacements, we have carried out this experiment for all the samples in order to further investigate the structural modification.

The recent studies have shown that Raman scattering technique can provide a very useful insight to the 'softening' of dynamic ferroelectric modes [29] and spin phonon coupling [30] in $ABO_3$ type perovskite. Theoretical group analysis predicts thirteen (13) zone-centre Raman-active optical phonon modes: $4A_1+9E$ for BFO at room temperature [31]. Where, $A_1$ modes are polarized along z-axis and E modes in x-y plane. By fitting the measured spectra and decomposing the fitted curves into individual lorentzian components, the peak position of each component, i.e. the natural frequency (cm$^{-1}$) of each Raman active mode, has been obtained for all the samples and the typical fitting for BFO has been shown in Fig. 4. Two sharp peaks at around 143 and 175 cm$^{-1}$ for BFO merged and results in the peak at 158 cm$^{-1}$ for the co-substituted samples which could be assigned as $A_1$-1 and Peak at around 213 cm$^{-1}$ as $A_1$-3 phonon modes respectively. E modes are assigned to other phonon modes located in the range of 50 – 750 cm$^{-1}$ as shown in Fig. 4. The frequency of the mode is proportional to $(k/M)^{1/2}$, where, k is the force constant and M is the reduced mass. The average mass of the A-sites significantly decreases with the increase in substitution concentration because atomic mass of Ca is about 81% less than that of Bi which leads to shift of $A_1$-1 and $A_1$-2 phonon modes to higher frequency side as the frequency of the phonon mode is inversely proportional to the reduced mass. The peak broadening and decrease in the intensity of low frequency phonon modes (such as $A_1$ modes) with the increase in co-substitution percentage indicates the chemical pressure-induced bond shortening and lattice distortion [32].

The increase in the relative intensity of E mode at 607 cm$^{-1}$ with the increase in substituent percentage also indicates the appearance of orthorhombic crystal symmetry (as shown in Fig. 4) [33]. The above results demonstrate the existence of the chemical pressure-induced rhombohedral–orthorhombic phase transition in BFO by co-substitution of Ca & Mn in Bi & Fe sites respectively. Similar results about the phase transition from the rhombohedral to orthorhombic symmetry in bismuth ferrite under high pressure using Mao-Bell diamond anvil cell have been observed by Y. Yang et al [34].

The FT-IR spectra of all the samples have been shown in Fig. 5 in the wave number range of 400-750 cm$^{-1}$. Typical band characteristics of metal oxygen bonds were observed in the range of 425-575 cm$^{-1}$. The absorption peaks at ~452 and ~554 cm$^{-1}$ are due to O-Fe-O bending of $FeO_6$ group and Fe-O bond stretching in the perovskite structure. These band positions are found to be in agreement with the characteristic infrared absorption bands of BFO [35] The gradual shift in the frequency of the Fe-O stretching mode indicates the formation of the solid solution. It indicates that substituent ions are at the corresponding sites in BFO and there is no nanoscale phase separation. However, absorption peak around 669 cm$^{-1}$ corresponds to the water vapor from atmosphere. With the increase in substituents percentage, the absorption peaks should shift to the higher wave number side up to 15 %, due to increased chemical pressure then to lower one for higher substitution (20 %) because the lattice strain relaxes with the appearance of orthorhombic crystal symmetry fraction (higher crystal symmetry). It is also consistent with the quantitative crystallographic phase contribution observed by double phase Rietveld analysis).

In order to find out the effect of chemical pressure on the band gap of BFO, we have recorded the diffuse reflectance spectra of all the samples. UV-Vis diffuse reflectance spectra were converted into absorption readings according to the Kubelka-Munk (K-M) method [36]. The absorption spectrum of the samples transformed from the diffuse reflection spectra using Kubelka-Munk function,

$$F(R) = \frac{(1-R)^2}{2R} \quad \text{------------ (2)}$$

Where, R is diffuse reflectance. Since BFO has a distorted cubic perovskite structure, there is a point group symmetry breaking from $O_h$ to $C_{3v}$ [37, 38]. There are expected six transitions between 0 and 3 eV by considering $C_{3v}$ local symmetry of Fe$^{3+}$ ions (High spin configuration $t_{2g}^3 e_g^2$) in BFO and using the correlation group and subgroup analysis for the symmetry breaking from $O_h$ to $C_{3v}$ [39]. In our case all six transitions were observed which lie in the range between 1.3 to 3 eV for BFO as shown in Fig. 6. However, these peaks vanish with the increase in the substitution concentration and finally there are only three peaks (low relative intensity) in BCFM-20. This indicates the structural transition from the distorted cubic perovskite (lower symmetry) to orthorhombic structure (higher symmetry) due to internal chemical pressure which arises as a result of size mismatch between substitution and host cations which corresponds to the modification in FeO$_6$ local environment. The analysis of XRD patterns, Raman spectroscopy and FTIR spectra (discussed above) also support well in this context.

The magnetization versus magnetic field (M-H) plots at room temperature of all samples with a maximum applied magnetic field of ± 80 kOe have been shown in Fig. 7 and magnetic parameters have been enlisted in table 1. The magnetic moment of Fe$^{3+}$ cations in BFO, are ferromagnetically coupled in pseudo cubic (111) planes but antiferromagnetically between adjacent planes and it is surrounded by six O$^{2-}$ ions in the common vertex of two adjacent FeO$_6$ octahedra. Magnetic hysteresis loop for x = 0.000 shows the characteristics of antiferromagnetic material. However, the co-substitution of Ca and Mn induces weak ferromagnetism. The unsaturated hysteresis loops and presence of small remnant magnetization reveals the presence of antiferromagnetic with weak ferromagnetism.

The tolerance factor of the substituted samples decreases with the increase in concentration of substituted elements because ionic size of $Ca^{2+}$ is smaller than that of $Bi^{3+}$. This leads to the increase in the octahedral tilt and the Fe-O-Fe bond angles which, in turn increases the superexchange interaction between the two antiferromagnetically aligned $Fe^{3+}$ cations with the suppression of cycloid spin structure [40]. The evolution of weak ferromagnetism in substituted samples is accounted due to the canting of antiferromagnetically ordered spins because of structural distortion as the there is no contribution of magnetization from the secondary phases as $Bi_2Fe_4O_9$ because it is paramagnetic at room temperature [41]. The maximum magnetization increases with increasing the substitution concentration up to 15%. But the enhancement in magnetization is small for low substituent concentration because the low substitution concentration cannot completely destroy the spin cycloid structure. However, at the morphological phase boundary (x = 0.15), the remnant and maximum magnetization at 8 T reaches a maximum which indicates almost broken spin cycloid structure. The further increase in substitution concentration results in the reduction of magnetization due to the appearance of complete antiferromagnetic ordering in the orthorhombic structure because of the significant contribution from the crystallographic phase of *Pbnm* space group (as obtained from the quantitative crystallographic phase contribution by double phase Rietveld analysis).

The substitution and other effects can break the cycloids in BFO, which has been reported to affect the magnetic properties of the samples, but presence of magnetic impurities ($Fe_2O_3$ or $Fe_3O_4$) may be the cause for the same. In the present samples, the presence of $Fe_2O_3$ magnetic impurity has not been detected form XRD (X-Ray diffraction) pattern analysis. Moreover, we have not observed ferromagnetic hysteresis loop for BFO which has largest amount of impurity phases among all samples. Also, the area of ferromagnetic hysteresis loop and coercivity increases with the increase in substitution percentage up to 15 % and then decreases sharply due to appearance of complete antiferromagnetic ordering in the orthorhombic structure. It also support our crystallographic phase percentage obtained from Rietveld analysis of XRD patterns which shows that the percentage of the orthorhombic crystal symmery (*Pbnm* space group) is dominant in the 20 % of co-substitution. This result can be attributed to the fact that the weak ferromagnetism in the substituted samples might be due to broken cycloid spin structure which leads to canting of the antiferromagnetic spin structure, which also deserves further study. These results indicates the morphological phase boundary near x = 0.15 which is consistent with the Rietveld analysis.

## 4. Conclusions

The co-substitution of Ca and Mn has resulted in structural transition from rhombohedral symmetry (space group *R3c*) to orthorhombic (*Pbnm*) as indicated by the XRD as well as Raman spectra analysis. Co-substitution with Ca and Mn significantly enhances the remnant magnetization as well as maximum magnetization at 80 kOe for the sample BCFM-15 (at morphotrophic phase boundary consistent with the Rietveld analysis) due to the broken cycloid spin structure caused by the distortion in the crystal lattice. The crystallographic phase percentage have been quantified by double phase Rietveld analysis of all XRD patterns which shows that the orthorhombic crystal symmetry (*Pbnm* space group) is dominant in the BCFM-20 which results in the sharp decrease of the $M_S$, $M_r$ and $H_C$ in this case due to appearance of complete antiferromagnetic ordering (in the orthorhombic crystal structure). UV-Visible absorption spectra analysis also supports the modification in local $FeO_6$ environment and structural transition. The enhanced magnetic properties of BFO will be interesting for magnetic field sensor applications.

**Acknowledgment**

The authors gratefully acknowledge Dr. Dhanbir Singh Rana and his research group at IISER Bhopal for extending the MPMS facility.

**References**

[1]  W. Eerenstein, N.D. Mathur, J. F. Scott, Multiferroic and magnetoelectric materials, Nature 442 (2006) 759-765.

[2]  S.W. Cheong, M. Mostovoy, Multiferroics: a magnetic twist for ferroelectricity, Nature Mater. 6 (2007) 13-20.

[3]  R. Ramesh, N.A. Spaldin, Multiferroics: progress and prospects in thin films, Nature Mater. 6 (2007) 21-29.

[4]   A. Singh, V. Pandey, R.K. Kotnala, D. Pandey, Direct Evidence for Multiferroic Magnetoelectric Coupling in $0.9BiFeO_3$–$0.1BaTiO_3$, Phys. Rev. Lett. 101 (2008) 247602-247606.

[5]   J. Wang, J.B. Neaton, H. Zheng, V. Nagarajan, S. B. Ogale, B. Liu, D. Viehland, V. Vaithyanathan, D.G. Schlom, U.V. Waghmare, N.A. Spaldin, K.M. Rabe, M. Wuttig, and R. Ramesh, Epitaxial $BiFeO_3$ Multiferroic Thin Film Heterostructures, Science 299 (2003) 1719-1722.

[6]   C. Ederer, N.A. Spaldin, Weak ferromagnetism and magnetoelectric coupling in bismuth ferrite, Phys. Rev. B 71 (2005) 060401-060405.

[7]   I. Dzyaloshinsky, A thermodynamic theory of "weak" ferromagnetism of antiferromagnetics, J. Phys. Chem. Solids 4 (1958) 241-255.

[8]   T. Moriya, Anisotropic Superexchange Interaction and Weak Ferromagnetism, Phys. Rev. 120 (1960) 91-98.

[9]   I. Sosnowska, T. Peterlin-Neumaier, E. Steichele, Spiral magnetic ordering in bismuth ferrite, J. Phys. C 15 (1982) 4835-4846.

[10]  A.V. Zalesskii, A.A. Frolov, T.A. Khimich, A.A. Bush, Composition-induced transition of spin-modulated structure into a uniform antiferromagnetic state in a $Bi_{1-x}La_xFeO_3$ system studied using $^{57}Fe$ NMR, Phys. Solid State 45 (2003) 141-145.

[11]  V.R. Palkar, D.C. Kundaliya, S.K. Malik, S. Bhattacharya, Magnetoelectricity at room temperature in the $Bi_{0.9-x}Tb_xLa_{0.1}FeO_3$ system, Phys. Rev. B 69 (2004) 212102-212105.

[12]  Q. Xu, H. Zai, D. Wu, Y.K. Tanga, M.X. Xu, The magnetic properties of $BiFeO_3$ and $Bi(Fe_{0.95}Zn_{0.05})O_3$, J. Alloys Compd. 485 (2009) 13-16.

[13]  J. Wei, D. Xue, C. Wu, Z. Li, Enhanced ferromagnetic properties of multiferroic $Bi_{1-x}Sr_xMn_{0.2}Fe_{0.8}O_3$ synthesized by sol–gel process, J. Alloys Compd. 453 (2008) 20-23.

[14]  Z.X. Cheng, A.H. Li, X.L. Wang, S.X. Dou, K. Ozawa, H. Kimura, S.J. Zhang, T.R. Shrout, Structure, ferroelectric properties, and magnetic properties of the La-doped bismuth ferrite, J. Appl. Phys. 103 (2008) 07E507-07E507-3.

[15]  G. Le Bras, D. Colson, A. Forget, N. Genand-Riondet, R. Tourbot, P. Bonville, Magnetization and magnetoelectric effect in $Bi_{1-x}La_xFeO_3$ ($0 \leq x \leq 0.15$), Phys. Rev. B 80 (2009) 134417-134423.

[16] S. Ghosh, S. Dasgupta, A. Sen, H.S. Maiti, Low-Temperature Synthesis of Nanosized Bismuth Ferrite by Soft Chemical Route, J. Am. Ceram. Soc. 88 (2005) 1349-1352.

[17] M.M. Kumar, V. R. palkar, K. Srinivas and S.V. Suryanarayana, Ferroelectricity in a pure $BiFeO_3$ ceramic, Appl. Phys. Lett. 76 (2000) 2764-2766.

[18] B. Ramachandran, A. Dixit, R. Naik, G. Lawes and M. S. R. Rao, Weak ferromagnetic ordering in Ca doped polycrystalline $BiFeO_3$, J. Appl. Phys. 111 (2012) 023910-023910-5.

[19] D. Kothari, V. R. Reddy, A. Gupta, V. Sathe, A. Banerjee, S. M. Gupta, A. M. Awasthi, Multiferroic properties of polycrystalline $Bi_{1-x}Ca_xFeO_3$, Appl. Phys. Lett. 91 (2007) 202505-202505-3.

[20] A. Ianculescu, F.P. Gheorghiu, P. Postolache, O. Oprea, L.Mitoseriu, The role of doping on the structural and functional properties of $BiFe_{1-x}Mn_xO_3$ magnetoelectric ceramics, J. Alloys Compd. 504, 420 (2010).

[21] Z.X. Cheng, X.L. Wang, Y. Du, S.X. Dou, A way to enhance the magnetic moment of multiferroic bismuth ferrite, J. Phys. D: Appl. Phys. 43 (2010) 242001-242006.

[22] V.R. Palkar, Darshan C. Kundaliya, and S. K. Malik, Effect of Mn substitution on magnetoelectric properties of bismuth ferrite system, J. Appl. Phys. 93, (2003) 4337-4339.

[23] S. Chauhan, M. Kumara, S. Chhokera, S. C. Katyal, H. Singh, M. Jewariya, K. L. Yadav Multiferroic, magnetoelectric and optical properties of Mn doped $BiFeO_3$ nanoparticles, Solid State Comm. 152 (2012) 525-529.

[24] J.Z. Huang, Y. Shen, M. Li and C.W. Nan, Structural transitions and enhanced ferroelectricity in Ca and Mn co-doped $BiFeO_3$ thin films, J. Appl. Phys., 110 (2011) 094106-094106-6.

[25] P. Kumar and M. Kar, Effect of structural transition on magnetic and optical properties of Ca and Ti co-substituted $BiFeO_3$ ceramics, J. Alloys. Compd. 584 (2014) 566.

[26] J. Rodriguez-Carvajal Laboratory, FULLPROF, a Rietveld and pattern matching and analysis programs version, Laboratoire Leon Brillouin, CEA-CNRS, France (2010).

[27] R. Palai, R. S. Katiyar, H. Schmid, P. Tissot, S. J. Clark, J. Robertson, S. A. T. Redfern, G. Catalan, and J.F. Scott, β phase and γ- β metal-insulator transition in multiferroic $BiFeO_3$, Phys. Rev. B 77 (2008) 014110-11.

[28] J.P. Patel, A. Sngh, D. Pandey, Nature of ferroelectric to paraelectric phase transition in multiferroic $0.8BiFeO_3$–$0.2Pb$ (Fe 1/2 Nb 1/2) $O_3$ ceramics, J. Appl. Phys. 107 (2010) 104115-104115-7.

[29] G. Burns and B.A. Scott, Lattice Modes in Ferroelectric Perovskites: $PbTiO_3$, Phys. Rev. B 7 (1973) 3008-3010.

[30] S.M. Cho, H.M. Jang and T.Y. Kim, Origin of Anomalous Line Shape of the Lowest-Frequency A1(TO) Phonon in $PbTiO_3$, Phys. Rev. B 64(1) (2003) 014103-014103-11.

[31] D. Kothari, V.R. Reddy, V.G. Sathe, A. Gupta, A. Banerjee, and A.M. Awasthi, Raman scattering study of polycrystalline magnetoelectric $BiFeO_3$, J. Magn. Magn. Mater. 320 (2008) 548-552.

[32] G.L. Yuan, S. W. Or and H. L. W. Chan, Reduced ferroelectric coercivity in multiferroic $Bi_{0.825}Nd_{0.175}FeO_3$ thin film, J. Appl. Phys. 101 (2007) 024106-024106-4.

[33] J.W. Lin, T. Tite, Y.H. Tang, C.S. Lue, Y.M. Chang and J.G. Lin, Correlation of spin and structure in doped bismuth ferrite nanoparticle, J. Appl. Phys. 111 (2012) 07D910-07D910-3.

[34] Y. Yang, L.G. Bai, K. Zhu, Y.L. Liu, S. Jiang, J. Liu, J. Chen and X.R. Xing, High pressure effects on $BiFeO_3$ studied by micro-Raman scattering, J. Phys.: condens. Matt. 21 (2009) 385901-385905.

[35] R.K. Mishra, D.K. Pradhan, R.N.P. Choudhary and A. Banerjee, Effect of yttrium on improvement of dielectric properties and magnetic switching behavior in $BiFeO_3$, J. Phys: Condens. Matter 20 (2008) 045218-045223.

[36] P. Kubelka and F.Z. Munk, F. Ein Beitrag zur Optik derFarbanstriche, Tech. Phys. 12 (1931) 593-601.

[37] K. Takahashi, N. Kida and M. Tonouchi, Terahertz Radiation by an Ultrafast Spontaneous Polarization Modulation of Multiferroic $BiFeO_3$ Thin Films, Phys. Rev. Lett. 96(11) (2006) 117402-117405.

[38] M. O. Ramirez, A. Kumar, S.A. Denev, N. J. Podraza, X. S. Xu, R. C. Rai, Y. H. Chu, J. Seide, L. W. Martin, S. Y. Yang, E. Saiz, J. F. Ihlefeld, S. Lee, J. Klug, S. W. Cheong, M. J. Bedzyk, O. Auciello, D. G. Schlom, R. Ramesh, J. Orenstein, J. L. Musfeldt, and V.

Gopalan, Magnon sidebands and spin-charge coupling in bismuth ferrite probed by nonlinear optical spectroscopy, Phys. Rev. B 79 (2009) 224106-224114.

[39] B. Ramachanran, A. Dixit, R. Naik, G. Lawes, M.S.R. Rao, Charge transfer and electronic transitions in polycrystalline $BiFeO_3$, Phys. Rev. B 82 (2010) 012102-012105.

[40] C. H. Yang, D. Kan, I. Takeeuchi, V. Nagarajan and J. Seidel, Doping $BiFeO_3$: approaches and enhanced functionality, Phys. Chem. Chem. Phys. 14 (2012) 15953-15962.

[41] N. Shamir, E. Gurewitz and H. Shaked, The magnetic structure of $Bi_2Fe_4O_9$ - analysis of neutron diffraction measurements, Acta Cryst., A34 (1978) 662-666.

**Table 1** Magnetic parameters of $Bi_{1-x}Ca_xFe_{1-x}Mn_xO_3$ (for x = 0.000 - 0.200) ceramics annealed at 700 °C. Where, $M_S$ = magnetization at maximum applied field, $H_C$ = Coercive field, emu/g = emu/gram at applied magnetic field of 80 kOe.

| $Bi_{1-x}Ca_xFe_{1-x}Mn_xO_3$ | Crystallite Size (nm) | $M_S$ at 80 kOe (emu/g) | $M_r$ (emu/g) | $H_C$ (kOe) |
|---|---|---|---|---|
| X = 0.000 | 95.6 | 0.6420 | 0.0086 | 0.1692 |
| X = 0.025 | 75.8 | 0.6682 | 0.0093 | 1.0694 |
| X = 0.050 | 47.4 | 0.8795 | 0.0507 | 3.7878 |
| X = 0.100 | 29.2 | 1.0044 | 0.1090 | 7.6293 |
| X = 0.150 | 24.8 | 1.1472 | 0.1237 | 7.9089 |
| X = 0.200 | 26.6 | 1.1426 | 0.0411 | 1.9393 |

**Table 2:** Lattice parameters and crystallographic phase contribution obtained by the Rietveld refinement of XRD patterns for $Bi_{1-x}Ca_xFe_{1-x}Mn_xO_3$ for x = 0.000 - 0.200.

| Space Group | *R3c* | *Pbnm* |
|---|---|---|
| **Value of a, b, c in (Å)** | | |
| X = 0.000 | a = b = 5.5805(2), c = 13.8536(1) | |
| X = 0.025 | a = b = 5.5628(5), c = 13.8136(4) | a = 5.5485(8), b = 5.5009(0), c = 7.9105(3) |
| X = 0.050 | a = b = 5.5573(8), c = 13.7921(1) | a = 5.5469(8), b = 5.5032(0), c = 7.9035(3) |
| X = 0.100 | a = b = 5.5550(1), c = 13.7636(0) | a = 5.5405(8), b = 5.5179(0), c = 7.8915(3) |
| X = 0.150 | a = b = 5.5407(0), c = 13.6733(6) | a = 5.5348(2), b = 5.5245(8), c = 7.8379(1) |
| X = 0.200 | a = b = 5.6126(4), c = 13.6125(8) | a = 5.4859(1), b = 5.5382(9), c = 7.8055(9) |
| **Crystallographic Phase contribution of all samples** | | |
| X = 0.000 | 100 % | 0 % |
| X = 0.025 | 95.96 % | 4.04 % |
| X = 0.050 | 93.83 % | 6.17 % |
| X = 0.100 | 84.43 % | 15.57 % |
| X = 0.150 | 56.64 % | 43.36 % |
| X = 0.200 | 30.52 % | 69.48 % |

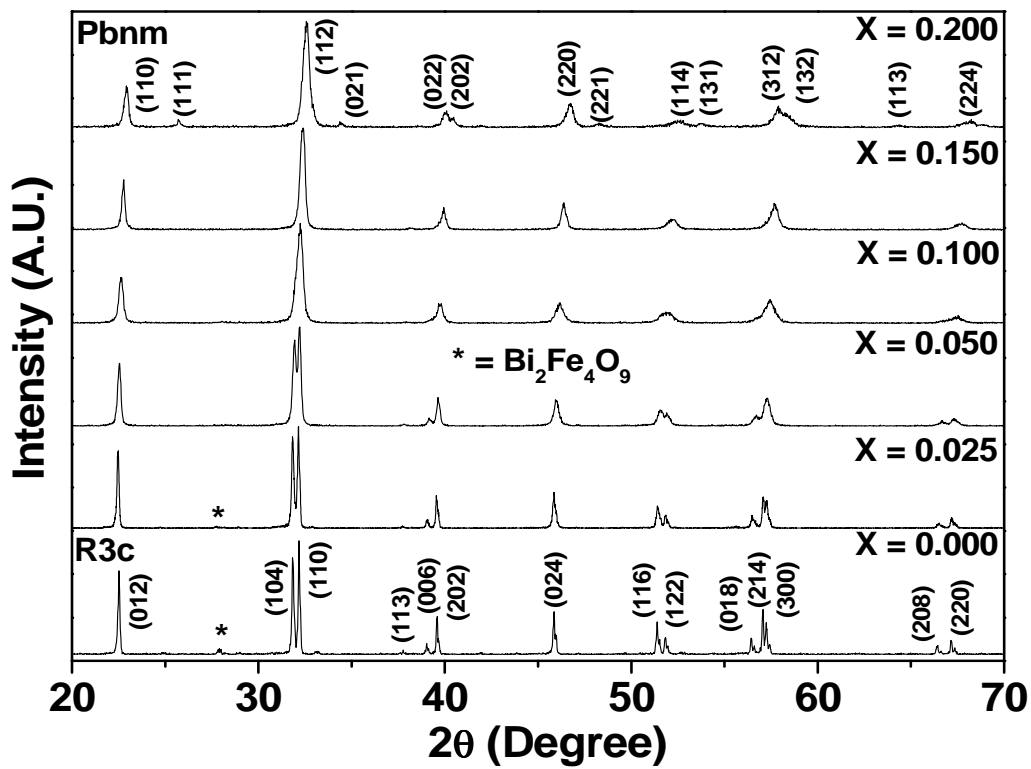

**Fig. 1.** XRD patterns of $Bi_{1-x}Ca_xFe_{1-x}Mn_xO_3$ samples for x = 0.000 - 0.200.

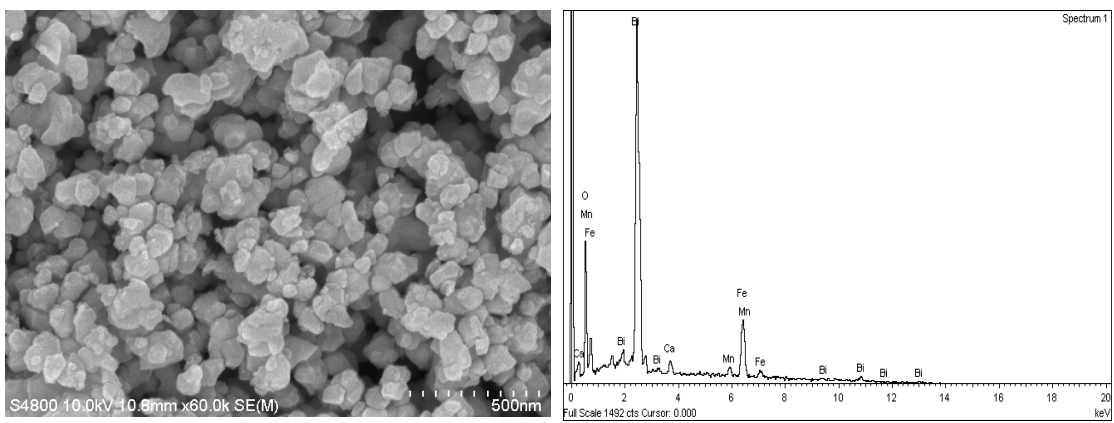

**Fig. 2.** FE-SEM image and EDS pattern of BCFM-05 sample.

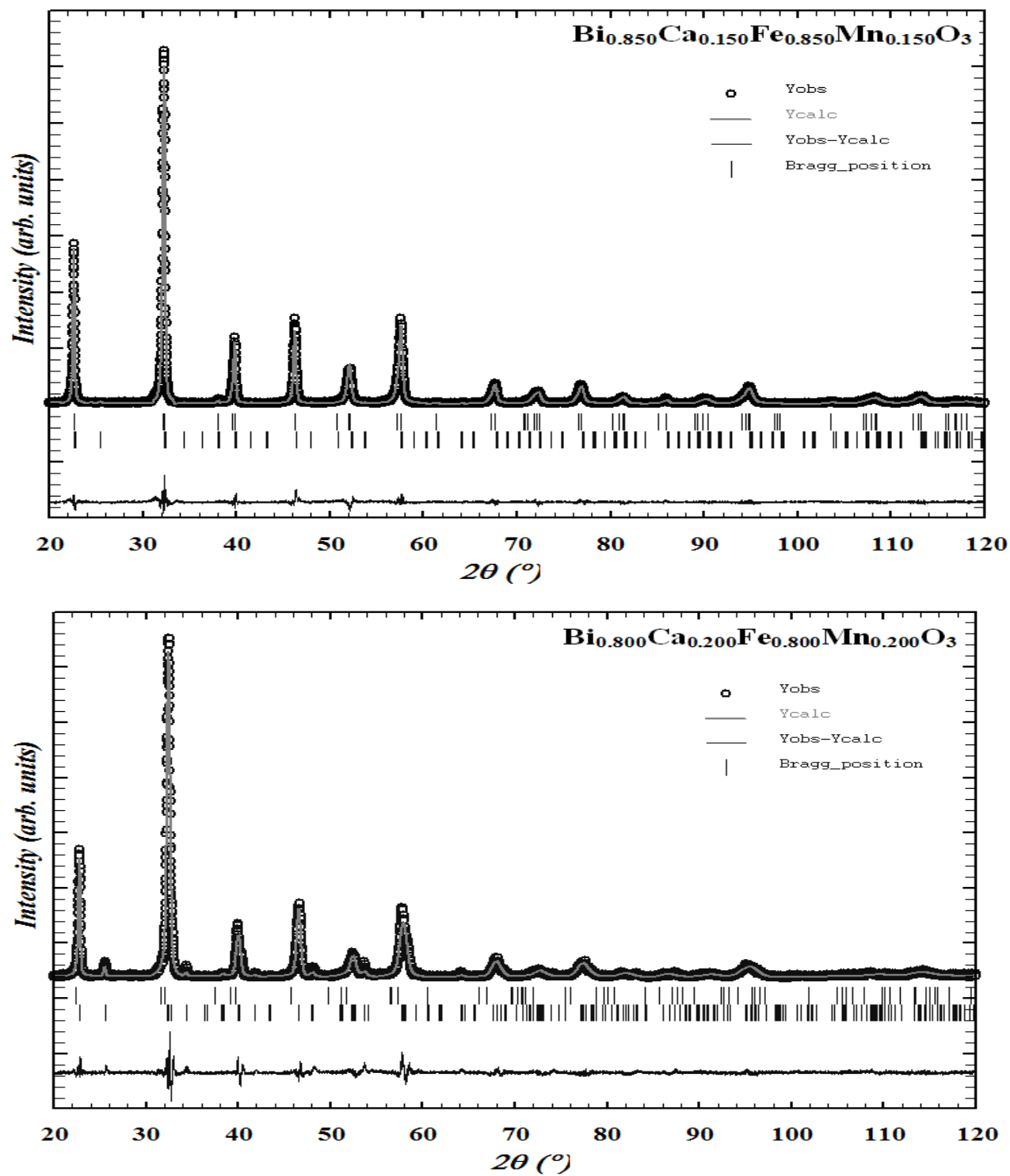

**Fig. 3.** Rietveld refined XRD patterns of $Bi_{1-x}Ca_xFe_{1-x}Mn_xO_3$ samples (x = 0.150 & 0.200).

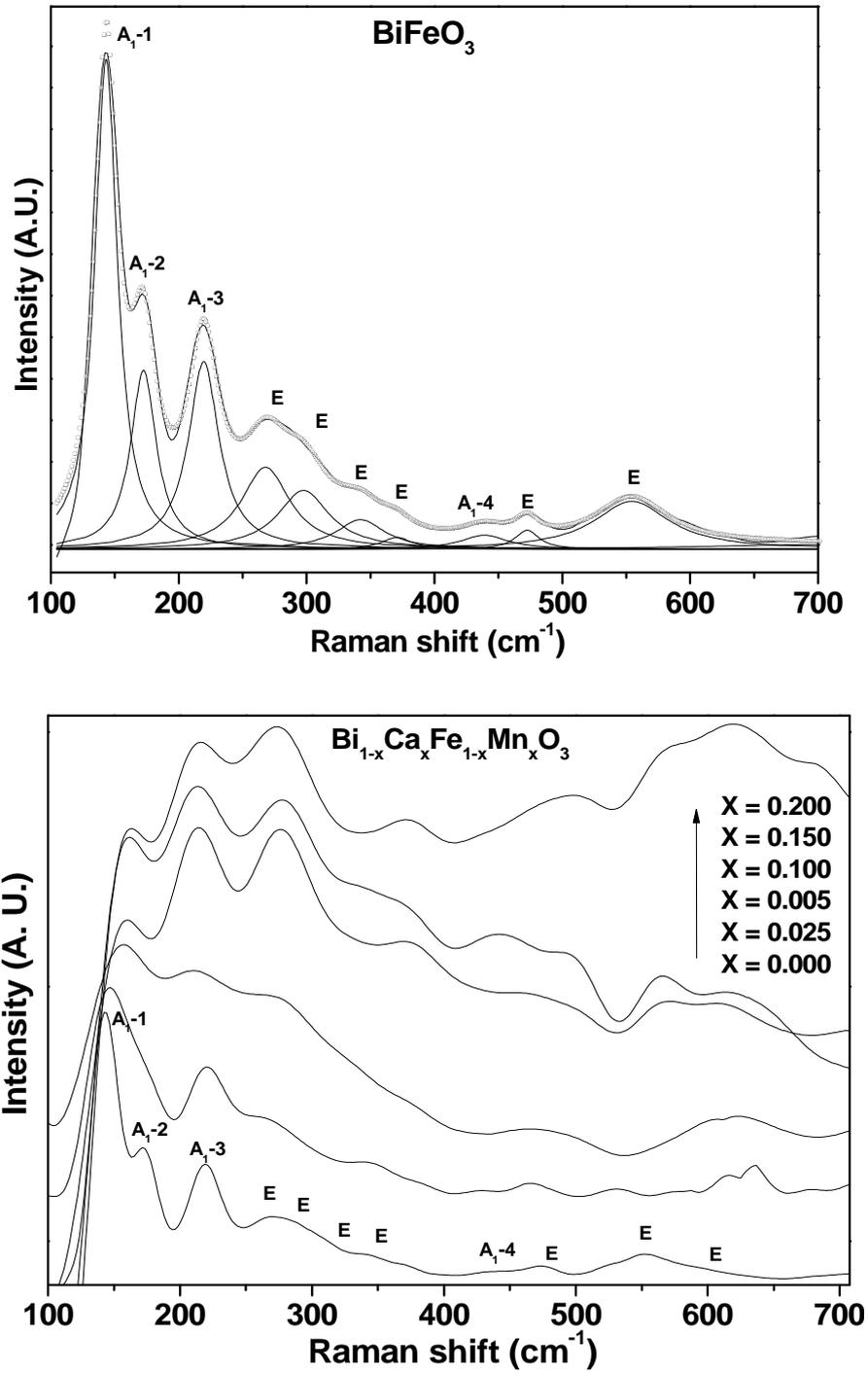

**Fig. 4.** Raman scattering spectrum of $BiFeO_3$ with its phonon modes deconvoluted into individual lorentzian components and spectra of $Bi_{1-x}Ca_xFe_{1-x}Mn_xO_3$ samples for x=0.000 - 0.200.

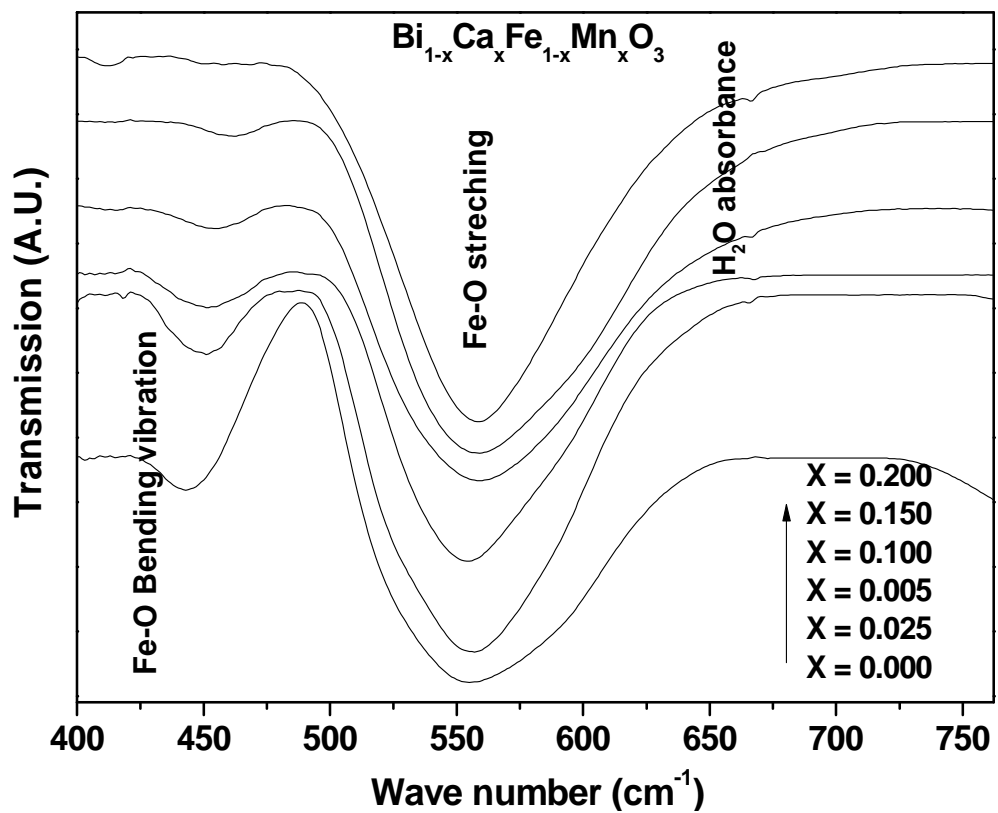

**Fig. 5.** FT-IR spectrum of $Bi_{1-x}Ca_xFe_{1-x}Mn_xO_3$ samples for x = 0.000 - 0.200.

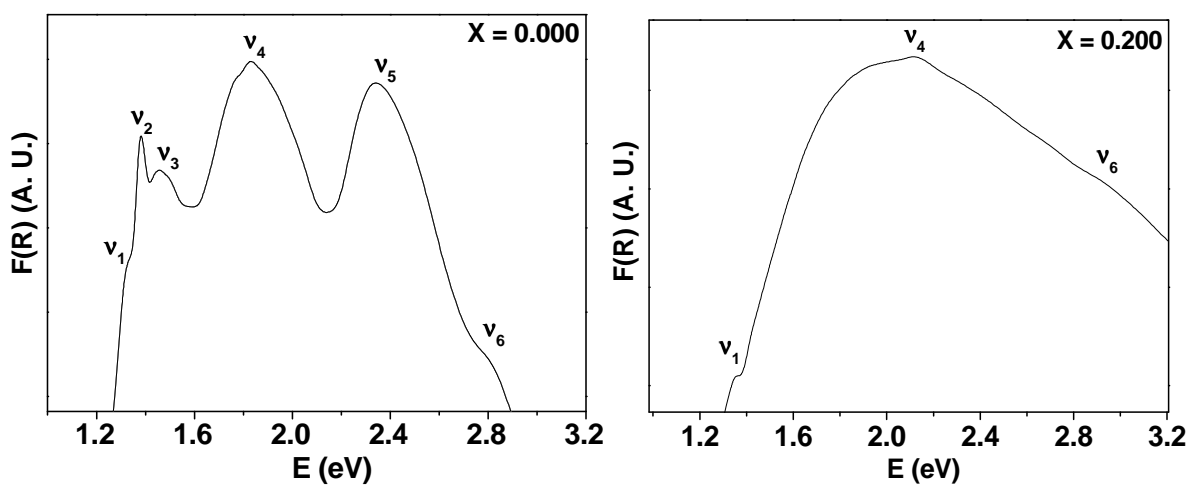

**Fig. 6**. UV-Visible absorption spectra of $Bi_{1-x}Ca_xFe_{1-x}Mn_xO_3$ samples for x = 0.000 & 0.200 in the 1 – 3.2 eV energy range.

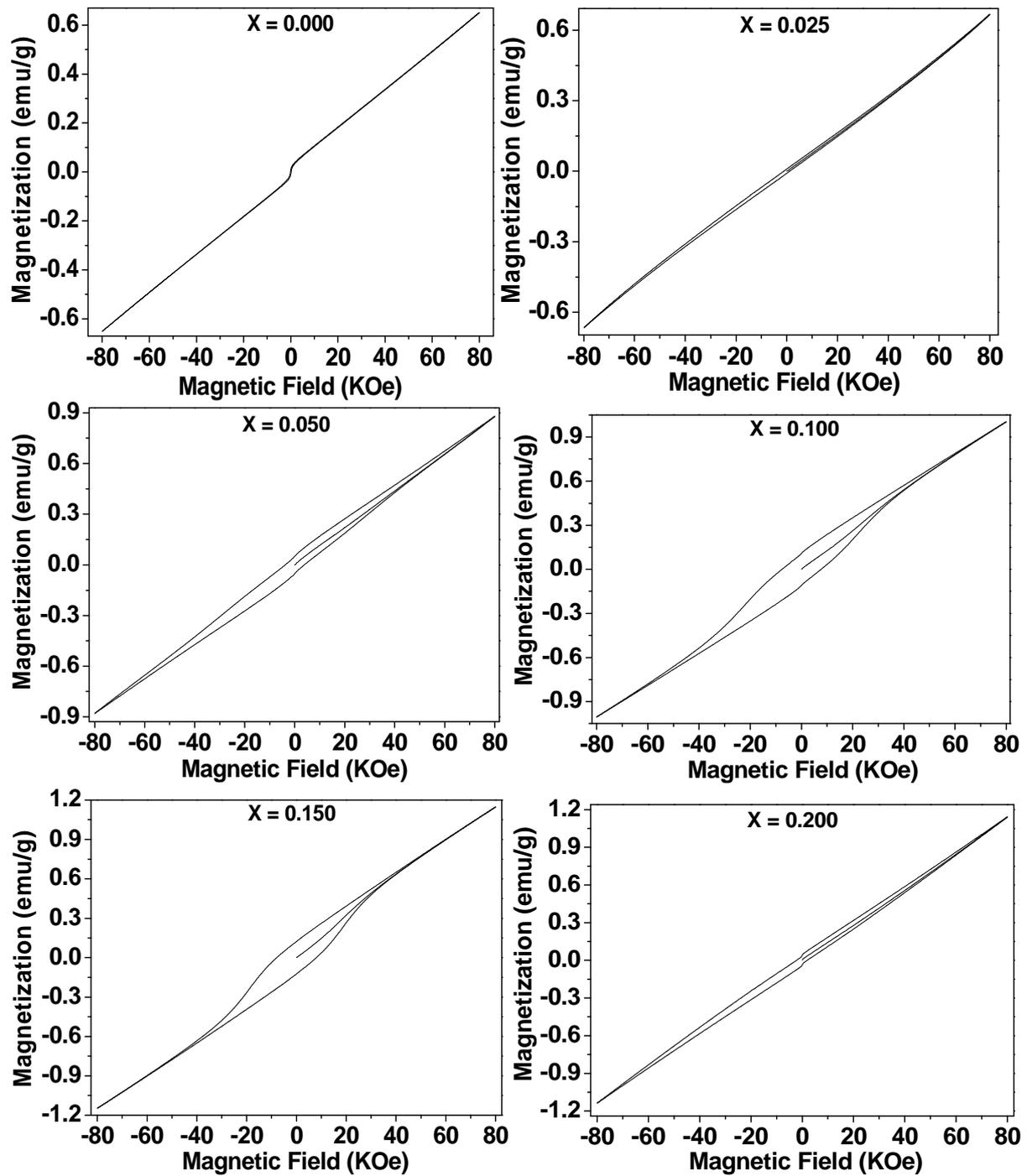

**Fig. 7**. M-H loops of $Bi_{1-x}Ca_xFe_{1-x}Mn_xO_3$ samples for x = 0.000 - 0.200.